\documentclass[10pt,conference]{IEEEtran}
\IEEEoverridecommandlockouts
\usepackage{amsmath,amssymb,amsfonts}
\usepackage{algorithmic}
\usepackage{algorithm}
\usepackage{array}
\usepackage{textcomp}
\usepackage{stfloats}
\usepackage{url}
\usepackage{verbatim}
\usepackage{graphicx}
\usepackage{cite}
\usepackage[colorlinks,linkcolor=blue,anchorcolor=blue,citecolor=blue,urlcolor=purple]{hyperref}
\usepackage{tcolorbox}
\usepackage{listings}
\usepackage{xcolor}
\usepackage[normalem]{ulem} 
\usepackage{graphicx}  
\usepackage{caption}
\usepackage{subcaption}
\usepackage{makecell}
\usepackage{color}
\usepackage{threeparttable}

\usepackage{float} 
\usepackage{pifont}
\usepackage{fontawesome}
\usepackage{multirow}
\usepackage{soul,framed}
\usepackage{cases}
\tcbuselibrary{most}
\usepackage{colortbl}
\usepackage{balance}

\newtcolorbox{mybox}[2][]
  {colback = white, colframe = black,
    colbacktitle = gray, enhanced,
    attach boxed title to top left={yshift=-2mm,xshift = 4mm},
    title=#2,#1}

\usepackage{multicol}
\usepackage{enumitem}
\usepackage{booktabs}
\definecolor{Black}{rgb}{0,0,0}
\definecolor{White}{rgb}{1,1,1}

\usepackage{ifthen}
\usepackage{fontawesome}
\newcolumntype{\CeX}{>{\centering\let\newline\\\arraybackslash}X}%
\newcommand{\TwoSymbolsAndText}[3]{%
 \begin{tabularx}{\textwidth}{c\CeX c}%
   #1 & #2 & #3
 \end{tabularx}%
}
\usepackage{minted}
\tcbuselibrary{listings}
\tcbuselibrary{minted}

\newtcbinputlisting[use counter=filePrg, number format={\arabic}]{\codeFromFile}[9]{ %
 listing engine=minted,
 minted language=#1,
 minted style=vs,
 listing file={#2},
 minted options={fontsize=#8,autogobble,#9,breaklines,firstline=#5,lastline=#6,highlightlines=#7, breakafter=.},
 listing only,
 size=title,
 arc= 0mm,
 breakable,
 enhanced jigsaw, 
 colframe=white,
 coltitle=White,
 boxrule=0mm,
 colback=white,
 coltext=Black,
 title=\TwoSymbolsAndText{\faCode}{%
   \textbf{Long Method \thetcbcounter}\ifthenelse{\equal{#3}{}}{}{\textbf{:} \textit{#3}}%
 }{\faCode},
 notitle,
 label=filePrg:#4,
}
\usepackage{cleveref}
\crefname{filePrg}{file program}{file programs}

\def\BibTeX{{\rm B\kern-.05em{\sc i\kern-.025em b}\kern-.08em
    T\kern-.1667em\lower.7ex\hbox{E}\kern-.125emX}}

\begin{document}


\title{Where Is Self-admitted Code Generated by Large Language Models on GitHub? 
\thanks{\textsuperscript{*}Corresponding author: Xing Hu, xinghu@zju.edu.cn}
}

\makeatletter
\def\@IEEEauthorrefmark#1{\textsuperscript{#1}}  
\makeatother

\author{
   \IEEEauthorblockN{
        Xiao Yu\textsuperscript{1},  
        Lei Liu\textsuperscript{2},  
        Xing Hu\textsuperscript{1*},  
        Jin Liu\textsuperscript{3},  
        and Xin Xia\textsuperscript{1}  
    }
\IEEEauthorblockA{
$^{1}$The State Key Laboratory of Blockchain and Data Security, Zhejiang University, Hangzhou, China, \\ xiao.yu@zju.edu.cn, xinghu@zju.edu.cn, xin.xia@acm.org \\
$^{2}$Faculty of Electronic and Information Engineering, Xi’an Jiaotong University, Xi’an, China, lei.liu@stu.xjtu.edu.cn \\
$^{3}$School of Computer Science, Wuhan University, Wuhan, China, jinliu@whu.edu.cn \\
}
}


\maketitle

\begin{abstract}
The increasing use of Large Language Models (LLMs) in software development has garnered significant attention from researchers evaluating the capabilities and limitations of LLMs for code generation. However, much of the research focuses on controlled datasets such as HumanEval, which do not adequately capture the characteristics of LLM-generated code in real-world development scenarios. To address this gap, our study investigates self-admitted code generated by LLMs on GitHub, specifically focusing on instances where developers in projects with over five stars acknowledge the use of LLMs to generate code through code comments. Our findings reveal several key insights: (1) ChatGPT and Copilot dominate code generation, with minimal contributions from other LLMs. (2) Projects containing ChatGPT/Copilot-generated code appears in small/medium-sized projects led by small teams, which are continuously evolving. (3) ChatGPT/Copilot-generated code generally is a minor project portion, primarily generating short/moderate-length, low-complexity snippets (e.g., algorithms and data structures code; text processing code).  (4) ChatGPT/Copilot-generated code generally undergoes minimal modifications, with bug-related changes ranging from 4\% to 12\%.  (5) Most code comments only state LLM use, while few include details like prompts, human edits, or code testing status. Based on these findings, we discuss the implications for researchers and practitioners.

\end{abstract}


\section{Introduction}

Large Language Models (LLMs), trained on massive amounts of text and code, have been widely applied to various software engineering tasks ~\cite{yang2024exploring, Yu2024fire, chen2022codet, zhao2024coding, Yu2024makes, li2025large}. 
Their growing adoption, particularly OpenAI's GPT series models, in software development has attracted considerable interest among researchers in examining the quality of code generated by these models. Some studies focus on proposing benchmarks for code generation, such as the widely used  HumanEval\cite{chen2021evaluating}, MBXP\cite{austin2021program}, as well as recently proposed 
CoderEval\cite{yu2024codereval}, ClassEval\cite{du2024evaluating}, and EvoCodeBench \cite{li2024evocodebench}, which address more complex real-world programming tasks (i.e., non-standalone function generation and class-level code generation). Additionally, other research endeavors aim to explore various quality attributes of the LLMs-generated code, such as code complexity \cite{liu2024no}, code security ~\cite{pearce2022asleep,perry2023users, siddiq2022empirical}, code maintainability ~\cite{yeticstiren2023evaluating}, and the error types of the generated code \cite{tambon2025bugs,dou2024s, zeng2024classifying}. 
These investigations offer valuable insights into LLMs' code generation capabilities and limitations. However, they predominantly rely on controlled datasets, which fail to fully capture the nuances of code generated by developers employing LLMs in practical scenarios.

To address the issue, we delve into the characteristics of self-admitted code generated by LLMs on GitHub, explicitly focusing on instances where developers acknowledge the use of LLMs to generate code through code comments. This self-admitted code reflects, to some extent, the characteristics of the code generated by developers using LLMs in real-world development scenarios.  
We begin by manually searching GitHub \cite{github_search} using keywords like ``\textit{generated by M}'', where \textit{M} represents commonly used LLMs. We specifically target the models included in the code generation leaderboards ~\cite{bigcode-models-leaderboard,code-generation-on-humaneval,leaderboard}, as these leaderboards compare the performance of various LLMs (e.g., GPT-4, DeepSeek, Qwen, CodeLlama, CodeGeeX) on various code generation benchmarks like HumanEval, MBPP, and MultiPL-E. 
We find that applications based on OpenAI's GPT models (i.e., ChatGPT and Copilot) are the most frequently utilized for generating code on GitHub. In contrast, other LLMs have either not been used to generate code or have only generated minimal code on GitHub. For example, searches for keywords such as ``\textit{generated by deepseek}'' across Python, Java, and JavaScript repositories yielded only 26 code snippets, all from repositories with fewer than 5 stars.
Therefore, our analysis in this paper focuses exclusively on the characteristics of code generated by ChatGPT and Copilot.
Next, we utilize the official GitHub REST API ~\cite{github_rest_api} to automatically identify and collect files/projects containing GPT-generated code \footnote{ChatGPT and Copilot are both based on OpenAI's GPT models. Thus, in the following text, we will use the term ``GPT-generated code'' to collectively refer to code generated by GPT-3.5, GPT-4.0, GPT-4o, ChatGPT, or Copilot.} in the common languages (Python, Java, C/C++, JavaScript, and TypeScript). 
Specifically, we first conduct keyword searches, such as ``\textit{generated by ChatGPT}'' and ``\textit{generated by Copilot}'', to locate GitHub code files that include such keywords, retaining only those files that contain GPT-generated code. We then use these files to obtain their respective projects and project attributes, such as stars, contributors, commits, and issues.  Ultimately, we identify 118 Python, 24 Java, 27 C/C++, 28 JavaScript, and 32 TypeScript projects, and corresponding 136 Python, 75 Java, 121 C/C++, 213 JavaScript, and 334 TypeScript files containing GPT-generated code. 
Additionally, we analyze the cyclomatic complexity and cognitive complexity of the GPT-generated code using the SonarQube tool ~\cite{campbell2013sonarqube}. We also collect code commit information to analyze the types of code changes in GPT-generated code. Furthermore, we analyze the comments on the GPT-generated code. Our study addresses several Research Questions (RQs):

\noindent\textbf{RQ1: What are the characteristics of the projects containing the self-admitted GPT-generated code?}
In the projects with the self-admitted GPT-generated code, there are more GPT-generated Python (140), TypeScript (248), and JavaScript (162) code snippets than Java (41) and C/C++ (75). Most projects are typically not large (the median number of code files in the projects ranging from 12 to 53 across the five programming languages). They are often led by small or medium-sized teams, with a median contributor count between 2 and 11. However, the projects generally undergo continuous development and improvement, with a median number of commits between 52 to 426 across the five languages. 

\noindent\textbf{RQ2: What are the characteristics of the self-admitted GPT-generated code? } 
Developers mainly utilize ChatGPT/Copilot to generate  algorithm and data structure code \footnote{This code is responsible for utilizing common data structures for processing non-text data, along with fundamental algorithms} for Python, Java, C/C++, JavaScript, and TypeScript, accounting for 22\%, 62\%, 63\%, 36\%, and 44\%, respectively.  In the case of Python, generating code for text processing is also common, comprising 26\% of the total.
The self-admitted GPT-generated code is primarily relatively short, with median Lines of Code (LOC) ranging from 10 to 63 across the five languages. Their complexity is generally low, with median cyclomatic complexity values from 1 to 3 and cognitive complexity between 0 and 4 across the five languages.

\noindent\textbf{RQ3: What are the characteristics of the code changes made to the self-admitted GPT-generated code? } 
When developers integrate the self-admitted GPT-generated code into GitHub projects, they may modify this code for various reasons over time. Upon investigation, we find that the GPT-generated code typically undergoes few modifications after its creation, with 8\% to 39\% of the code being changed across the five languages. The median number of added LOC ranges from 0 to 4 across the five languages, and the median number of deleted lines is 0, suggesting that only incremental adjustments are often necessary rather than extensive overhauls. Furthermore, only 4\% to 12\% of these modifications are due to bugs, indicating that the initial code quality is relatively high. The main types of bugs in the self-admitted GPT-generated code include algorithm logic errors, interface and dependency management issues, and grammar and semantic errors. Additionally, the most common reasons for non-bug-fix code changes are feature addition and refactoring.

\noindent\textbf{RQ4: How do the self-admitted GPT-generated code and code without explicit GPT-generation annotation differ within the same projects?} 
The median proportion of the self-admitted GPT-generated LOC to the total LOC in the projects varies from 0.02 to 0.03 across the five programming languages, indicating that the self-admitted GPT-generated code constitutes only a very small portion of the projects. Compared to code without explicit GPT-generation annotation \footnote{Code without explicit GPT-generation annotation likely consists mostly of manually written code, with a smaller portion that was generated by GPT but not self-admitted by developers.} with frequent changes, the self-admitted GPT-generated code undergoes very few changes or bug fixes.

\noindent\textbf{RQ5: How do developers annotate GPT-generated code?} Most code comments only simply state that the code is generated by ChatGPT/Copilot without offering any additional information.  A minority of comments detail the version of LLMs used (average 23\%), e.g., ChatGPT-3.5, Copilot, the timestamp of code generation (average 9\%), the prompt provided to ChatGPT/Copilot for code generation (average 11\%), the proportion of code generated by ChatGPT/Copilot in the snippet (average 4\%), whether the generated code undergoes human modifications (average 6\%), and whether further testing or reviews are necessary or completed (average 3\%). 

Our paper makes the following contributions: 

(1) To the best of our knowledge, we conduct the first comprehensive empirical study examining the characteristics of self-admitted GPT-generated code and its projects on GitHub.

(2) We present our findings and offer implications for practitioners and researchers regarding the adoption and development of LLMs for code generation in practice.


\section{Methodology}
\label{sec:methodology}

\subsection{Data Collection}
\label{sec: data collection}

We utilize the GitHub REST API 
to automatically collect projects containing the self-admitted GPT-generated code.

\noindent \textbf{\textit{Keywords.}} We initially collect code comments from GitHub containing tokens such as ``ChatGPT'', ``Copilot'', ``GPT3'', ``GPT4'', ``GPT-3'', and ``GPT-4''.  
Then, we analyze the most frequently occurring triplets, which include ``ChatGPT'', ``Copilot'', ``GPT3'', ``GPT4'', ``GPT-3'', and ``GPT-4''. We observe that triplets indicating code generated by ChatGPT/Copilot typically adhere to the pattern \textit{x}+\textit{y}+\textit{z}, where \textit{x} $\in$ \{generated, written, created, implemented, authored, coded\}, \textit{y} $\in$ \{by, through, using, via, with\}, and \textit{z} $\in$ \{ChatGPT, Copilot, GPT3, GPT4, GPT-3, and GPT-4\}. Thus, we employ these triplets, such as ``\textit{generated by ChatGPT}'', as keywords to locate the self-admitted GPT-generated code on GitHub — the GitHub search box does not differentiate between the letter case, and utilizing tokens such as \textit{generated by ChatGPT}'' also allows us to search for code comments containing variations like ``generated by ChatGPT-3.5''. We collect the GPT-generated code from GitHub continuously until April 2025.

\noindent \textbf{\textit{Retrieving code files.}}  
We conduct keyword searches, such as ``\textit{generated by ChatGPT}'', to retrieve code files containing such the keywords and then proceed to download these code files.
Following that, we invite three software developers with over five years of programming experience to review these files and determine whether they indeed indicate the presence of GPT-generated code, and to identify the specific lines within the code files that are generated by ChatGPT/Copilot. 
This is necessary because some developers may utilize ChatGPT/Copilot to generate other items, such as code comments (e.g., the code comment ``Description automatically generated by ChatGPT'' in the code file 
~\cite{packtools-commit-32ad} 
on GitHub), rather than source code. We also filter out repositories with GPT-generated code that have fewer than 5 stars, and the three developers manually review these to mitigate the impact of toy programs. 
Initially, we select the top 10 programming languages 
~\cite{Top10Language}
from GitHub to download code in these programming languages, but we find that the number of GPT-generated code snippets in C\#, PHP, Shell, and Ruby is less than 40, which may lead to statistically unreliable results. Consequently, we focus our analysis solely on GPT-generated code snippets in Python, Java, C/C++, JavaScript, and TypeScript on GitHub.
From an initial collection of 231 Python, 87 Java, 121 C/C++, 233 JavaScript, and 334 TypeScript code files, we filter out files that do not contain GPT-generated code and ultimately obtain 136 Python, 41 Java, 75 C/C++, 152 JavaScript, and 248 TypeScript files containing the GPT-generated code.

\noindent \textbf{\textit{Retrieving GitHub projects.}} We utilize the metadata of the collected code files to get their corresponding projects. Consequently, we obtain 118 Python, 24 Java, 27 C/C++, 28 JavaScript, and 32 TypeScript projects (as indicated in column \#Proj$_G$ in Table \ref{tab:project}).

\noindent \textbf{\textit{Retrieving code changes.}}
We leverage the project name and code file name as unique identifiers to retrieve the complete commit history for each code file. Based on the unique hash value of each commit, we utilize the GitHub REST API
to obtain the content of the code file for its each code commit (to analyze the types of code changes in RQ3). Additionally, we record the timestamp of the first code commit for each file (i.e., when the code file containing GPT-generated code was initially uploaded to GitHub).

\noindent \textbf{\textit{Downloading code files and projects.}}
We download projects containing GPT-generated code in April 2025. The project characteristic analysis for RQ1 is based on these projects at that time.
We download the file containing GPT-generated code that was initially uploaded to GitHub. The characteristic analysis of GPT-generated code in RQ2 is based on the files at this timestamp, because GPT-generated code snippets may undergo subsequent human modifications. Therefore, we analyze the information pertaining to the GPT-generated code snippets at the time of their initial upload to GitHub.

\begin{figure*}[!ht]
    \centering
    \includegraphics[width=1\textwidth]{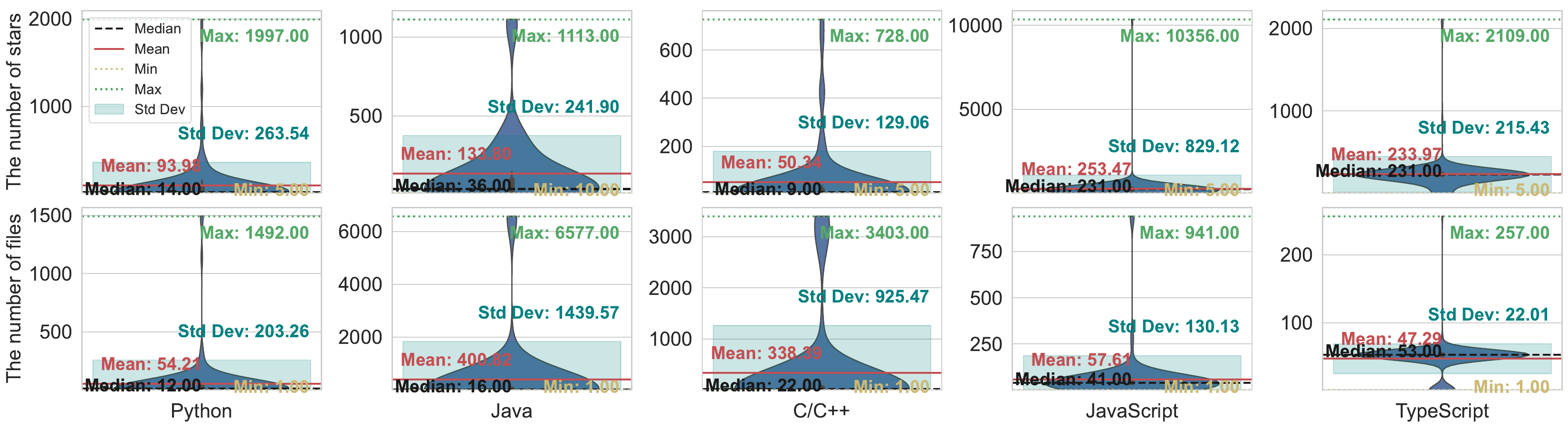}
    \vspace{-0.6cm}
    \caption{The distribution of the stars and files in the projects containing the GPT-generated code. }
    \vspace{-0.3cm}
\label{fig:project_metrics}
\end{figure*}

\subsection{Analysis}

\noindent\textbf{\textit{Code categorization.}} 
To systematically classify self-admitted GPT-generated code into distinct functional categories, we adopt a manual classification approach involving multiple human developers. The classification process was divided into two stages.
In the first stage, we randomly sample 40\% of the code for the pilot annotation to develop an initial taxonomy.  Two developers independently annotate the function types of the self-admitted GPT-generated code. During this process, the two developers carefully review the complete context of each code snippet. For cases of disagreement, the third developer facilitates structured discussions to reach consensus, finalizing the taxonomy before proceeding to full annotation.
In the second stage, we formally annotate the remaining 60\% of the self-admitted GPT-generated code. Two developers independently label each code snippet in a single round, and inter-rater reliability is measured using Cohen’s Kappa \cite{cohen1960coefficient}, resulting in a score of 0.83, indicating substantial agreement. All remaining discrepancies are resolved by discussion with the third developer. This process yields 15 mutually exclusive functional categories for self-admitted GPT-generated code (Table \ref{tab:codetype}).

\noindent \textbf{\textit{Code change categorization.}}  Following Ray et al. ~\cite{ray2014large} and Mockus et al. ~\cite{mockus2000identifying}, we identify commit logs that contain keywords such as ``error'', ``bug'', ``fix'', ``issue'', ``mistake'', ``incorrect'', ``fault'', ``defect'', ``flaw'', and ``vulnerability'' as bug-fix commits. When a developer submits a bug-fix commit, we observe whether there are code changes in the self-admitted GPT-generated code at the time of this commit. If code changes are detected, we task the three developers to examine whether the bug-fix information in the commit log corresponds to a bug-fix applied to the GPT-generated code. If so, it indicates that the GPT-generated code contained a bug previously. 
Subsequently, we request that the three developers identify the type of bug. For code changes made for reasons other than bug fixes, we also engage them to analyze the types of code changes. 
Specifically, similar to the code categorization process mentioned earlier, two developers first conduct pilot annotation on 40\% of the changes to build a preliminary classification system. A third developer reviews all cases with disagreements and organizes consensus discussions to determine the final classification system.
Then, the two developers perform formal independent single-round annotation on the remaining 60\% of the changes. The Cohen’s Kappa coefficients for bug categorization and change type categorization are 0.79 and 0.77, respectively, indicating high inter-rater reliability. This process ultimately results in 6 categories of bug-fix code changes (Table \ref{tab:bugfixtype}) and 6 categories of non-bug-fix code changes (Table \ref{tab:nonbugfixtype}).


\noindent \textbf{\textit{Code analysis.}} Following Bogner et al. \cite{bogner2022type} and Liu et al. \cite{liu2024no}, we utilize the static analysis tool SonarQube ~\cite{campbell2013sonarqube} to analyze GPT-generated code and its corresponding projects. We extract metrics such as LOC, number of files/classes/methods, cyclomatic complexity \cite{mccabe1976complexity}, and cognitive complexity \cite{campbell2018cognitive, cognitive-complexity}, which are commonly used to evaluate code complexity.
Cyclomatic complexity is the number of linearly independent paths through source code, determining how difficult it is to test.
Cognitive complexity evaluates the difficulty in understanding and reasoning about code from a human perspective. 

\section{Results}
\label{sec: results}

\subsection{ RQ1: The characteristics of the projects containing the self-admitted GPT-generated code }


\begin{table*}[!ht]
\centering
 \caption{The information of projects containing the self-admitted GPT-generated code.
     \vspace{-0.2cm}
     }
     \label{tab:project}
\fontsize{7.5pt}{8pt}\selectfont
\begin{tabular}{@{}l| l| l | l |l  |l  |l  |l  |l | l |l | l @{}}
\toprule

  ~& \#Proj$_{G}$ & \#Files$_P$ & \#LOC$_P$  & Stars &  Files  & Contributors & Commits & Commits$_{Fix}$ & Issues & \#Files$_G$ & \#Sni$_G$  \\ \midrule  

Python & 118 & 6.12K & 0.80M & 94,14 \tiny \makecell{\textcolor{blue}{2.0K} \\ \textcolor{red}{5}} & 54, 12 \tiny \makecell{\textcolor{blue}{1.5K} \\ \textcolor{red}{1}} &   8, 4 \tiny \makecell{\textcolor{blue}{222} \\ \textcolor{red}{2}}  &  292, 111 \tiny \makecell{\textcolor{blue}{2.9K} \\ \textcolor{red}{1}} & 43, 9 \tiny \makecell{\textcolor{blue}{853} \\ \textcolor{red}{0}}  & 8, 0 \tiny \makecell{\textcolor{blue}{199} \\ \textcolor{red}{0}}  & 136 & 140 \\ \midrule 

Java & 24 & 15.97K  & 2.14M & 134, 36 \tiny \makecell{\textcolor{blue}{1.1K} \\ \textcolor{red}{10}} & 401, 16 \tiny \makecell{\textcolor{blue}{6.6K} \\ \textcolor{red}{1}}  & 43, 3 \tiny \makecell{\textcolor{blue}{307} \\ \textcolor{red}{2}}  & 3986, 332 \tiny \makecell{\textcolor{blue}{54.1K} \\ \textcolor{red}{7}}  &  1019, 25 \tiny \makecell{\textcolor{blue}{12.1K} \\ \textcolor{red}{0}} &  18, 2 \tiny \makecell{\textcolor{blue}{302} \\ \textcolor{red}{0}} & 41 & 41   \\ \midrule  

C/C++ & 27 & 24.11K & 5.18M & 50, 9 \tiny \makecell{\textcolor{blue}{728} \\ \textcolor{red}{5}} & 338, 22 \tiny \makecell{\textcolor{blue}{3.4K} \\ \textcolor{red}{1}}  & 16, 11 \tiny \makecell{\textcolor{blue}{35} \\ \textcolor{red}{2}}  & 340, 52 \tiny \makecell{\textcolor{blue}{2.8K} \\ \textcolor{red}{9}}  &  69, 3 \tiny \makecell{\textcolor{blue}{778} \\ \textcolor{red}{0}} &  3, 0 \tiny \makecell{\textcolor{blue}{23} \\ \textcolor{red}{0}} & 75 & 75   \\ \midrule  

JavaScript & 28 & 3.92K & 2.90M & 253, 231 \tiny \makecell{\textcolor{blue}{10.4K} \\ \textcolor{red}{5}} & 58, 41 \tiny \makecell{\textcolor{blue}{941} \\ \textcolor{red}{1}}  & 3, 2 \tiny \makecell{\textcolor{blue}{61} \\ \textcolor{red}{2}}  & 373, 426 \tiny \makecell{\textcolor{blue}{858} \\ \textcolor{red}{5}}  &  51, 55 \tiny \makecell{\textcolor{blue}{230} \\ \textcolor{red}{0}} &  8, 6 \tiny \makecell{\textcolor{blue}{330} \\ \textcolor{red}{0}} & 152 & 162   \\ \midrule  

TypeScript & 32 & 11.4K & 3.41M & 234, 231\tiny \makecell{\textcolor{blue}{2.1K} \\ \textcolor{red}{5}} & 47, 53\tiny \makecell{\textcolor{blue}{257} \\ \textcolor{red}{1}}  & 3, 2\tiny \makecell{\textcolor{blue}{52} \\ \textcolor{red}{2}}  & 425, 426\tiny \makecell{\textcolor{blue}{7.3K} \\ \textcolor{red}{1}}  &  64, 55\tiny \makecell{\textcolor{blue}{2.3K} \\ \textcolor{red}{0}} &  7, 6\tiny \makecell{\textcolor{blue}{175} \\ \textcolor{red}{0}} & 248 & 248   \\

\bottomrule

\end{tabular}
\vspace{-0.2cm}
\end{table*}

\begin{table*}[!ht]
\centering
 \caption{The code type of the self-admitted GPT-generated code (the most frequent three types are highlighted in bold). \vspace{-0.2cm}
     }
    \label{tab:codetype}
\fontsize{8pt}{8pt}\selectfont
\begin{tabular}{@{}l | p{0.65cm} | p{0.65cm} | p{0.65cm} | p{0.65cm} | p{0.65cm} | p{0.65cm} | p{0.65cm} | p{0.65cm} | p{0.65cm} | p{0.75cm} | p{0.75cm} | p{0.75cm} | p{0.75cm} | p{0.75cm} | p{0.75cm}@{}}
\toprule
 ~ & Type1 &Type2 & Type3 & Type4 & Type5 & Type6 & Type7 & Type8 & Type9 & Type10 & Type11  & Type12 & Type13 & Type14 & Others\\
\midrule
Python & \textbf{26\%}& 4\%  & 11\% & \textbf{22\%} &   13\%&  7\%&  2\%&  \textbf{22\%}& 14\%& 4\%&  1\%&  9\%& 0\%& 1\%& 0\% \\
\midrule

Java &   \textbf{27\%}&   7\%&  \textbf{15\%}&   \textbf{62\%}&  7\%&  2\%& 2\%& 5\%& 15\%& 5\% & 7\%& 13\%& 2\%& 0\%& 0\%\\
\midrule

C/C++ &  12\%&   3\%&   5\%&   \textbf{63\%}&  7\%&  4\%& 8\%& 13\%& 0\%& 13\%& \textbf{19\%}& \textbf{17\%}& 0\%& 0\%& 0\%\\
\midrule

JavaScript &  9\% &   8\%&   7\%&   \textbf{36\%}&  10\%& 11\%& 1\%&  6\%& 0.7\%& 9\%& 0.7\%& 0.7\% & \textbf{13\%}& \textbf{20\%}& 0.7\%\\
\midrule

TypeScript &   \textbf{9\%}&   3\%&  \textbf{17\%}&  \textbf{44\%}&  10\%& 5\%& 8\%&  4\%& 0.4\%& 2\%& 0.4\%& 0\%& 0.4\%& 8\%& 0\%\\
\midrule
Avg& \textbf{14\%}& 5\%& \textbf{12\%}& \textbf{41\%}& 10\%& 6\%& 5\%& 9\%& 4\%& 6\%& 3\%& 5\%& 3\%&8\%& 0.2\%\\

\bottomrule
\end{tabular}
\begin{tablenotes}[flushleft]
\item \textbf{Type Explanation:} Type1 (Text processing), Type2 (Configuration and deployment), Type3 (Program input code, e.g., variable assignments and regular expressions), Type4 (Algorithm and data structure), Type5 (Network requests and API call), Type6 (User interface), Type7 (Testing and debugging), Type8 (File operations and I/O operations), Type9 (Machine learning and deep learning), Type10 (Framework and library), Type11 (Game development), Type12 (Scientific computation and numerical analysis), Type13 (Concurrency and multithreading), Type14 (Security), and Others (e.g., code for manipulating audio data, SQL queries, etc.). Note: Since the code may belong to multiple categories, the sum of each row does not equal 1.
\end{tablenotes}
\vspace{-0.5cm}
\end{table*}

Table \ref{tab:project} presents the characteristics of the projects containing the self-admitted GPT-generated code. ``\#Proj$_{G}$'', ``\#File$_{G}$'', and ``\#Sni$_{G}$''  represent the number of projects containing GPT-generated code, files containing GPT-generated code, and GPT-generated code snippets, respectively. ``\#Files$_{P}$'' and ``\#LOC$_{P}$'' denote the total number of files and LOC (including both self-admitted GPT-generated and code without explicit GPT-generation annotation) within these projects.  
The entries in the columns of ``Stars'', ``Files'', ``Contributors'', ``Commits'', ``Commits$_{fix}$'', and ``Issues'' represent the average, median, maximum, and minimum values for these metrics across all projects. For example, the entry $94, 14 \tiny \begin{array}{c} \textcolor{blue}{2.0K} \\ \textcolor{red}{5} \end{array}$ represents that, for these 118 Python projects, the average star count is 94, the median is 14, the maximum reaches 2.0K, and the minimum is 5.  
Due to space limitations, and considering that stars and file numbers are widely used to illustrate project popularity and size, Figure \ref{fig:project_metrics} only presents their distribution.

The projects containing the self-admitted GPT-generated code typically exhibit a relatively high median star count, ranging from 9 to 231, as well as file counts ranging from 12 to 53 across the five programming languages.
Although there are exceptional cases where projects reach as high as 2.0K stars and 1.5K files, most projects containing the self-admitted GPT-generated code are relatively not large. 
The median number of contributors varies between 2 and 11 across the five programming languages, with the majority of projects being developed by small to medium-sized  teams. The median number of commits ranges from 52 to 426, indicating that they are not static but undergo continuous development and improvement. 
However, the projects exhibit a relatively low frequency of bug-fix commits and issues, with median values ranging from 3 to 55 for bug-fix commits and 0 to 6 for issues. 

There are more projects containing the self-admitted GPT-generated Python code compared to other programming languages. However, in some JavaScript and TypeScript projects, e.g., \cite{gptlint, Bing_Chat_History}, there are multiple GPT-generated code files, resulting in a higher number of files containing JavaScript and TypeScript code than other languages. Overall, developers are more inclined to use ChatGPT/Copilot to generate Python, JavaScript, and TypeScript code. The primary reason for this preference is that Python is a widely used programming language, and its simple and readable syntax makes it easier for ChatGPT/Copilot to generate accurate Python code. Additionally, JavaScript and TypeScript are commonly used in business projects, where ChatGPT/Copilot is often employed to automate the creation of templated or repetitive code.


\begin{center}
\vspace{-15pt}
    \resizebox{\linewidth}{!}{
\begin{tabular}{l!{\vrule width 1pt}p{1\columnwidth}}
    \makecell{{\LARGE \faLightbulbO}}  &\textbf{Finding 1.} Most GitHub projects that contain the self-admitted GPT-generated code are relatively not large, typically developed by small to medium-sized teams. However, these projects are generally dynamic, undergoing continuous development and improvement. Developers primarily use ChatGPT and Copilot to generate code in Python, TypeScript, and JavaScript. 
\end{tabular}}
\vspace{-8pt}
\end{center}

\subsection{RQ2: The characteristics of the self-admitted GPT-generated code }

\begin{table*}[!ht]
\centering
 \caption{The number of files, classes, methods, and statements (code granularity) that ChatGPT/Copilot generates, and the information of the LOC, cyclomatic complexity, and cognitive complexity of the files, classes, methods, and statements. \vspace{-0.2cm}
     }
    \label{tab:code}
\fontsize{7.4pt}{8pt}\selectfont
\begin{tabular}{@{}p{1.0cm} | p{0.5cm} | p{0.55cm} | p{0.75cm} | p{0.45cm} | l | l | l | l | l | l | l | l | l | l@{}}
\toprule
 ~ & \#Files & \#Class & \#Method & \#Stat & LOC$_F$ & CC$_F$ & CogC$_F$ & LOC$_C$ & CC$_C$ & CogC$_C$ & LOC$_M$  & CC$_M$  & CogC$_M$ & LOC$_S$   \\
\midrule
Python & 51 &  7 & 49  & 29 &  123, 64\tiny \makecell{\textcolor{blue}{1.2K} \\ \textcolor{red}{1}} & 4,  3\tiny \makecell{\textcolor{blue}{26} \\ \textcolor{red}{0}} & 5, 3\tiny \makecell{\textcolor{blue}{49} \\ \textcolor{red}{0}} & 88, 26\tiny \makecell{\textcolor{blue}{455} \\ \textcolor{red}{11}} & 3, 3\tiny \makecell{\textcolor{blue}{5} \\ \textcolor{red}{2}} & 4,  3\tiny \makecell{\textcolor{blue}{9} \\ \textcolor{red}{1}} & 37, 20\tiny \makecell{\textcolor{blue}{406} \\ \textcolor{red}{5}} & 4,  3\tiny \makecell{\textcolor{blue}{26} \\ \textcolor{red}{0}} & 6, 4\tiny \makecell{\textcolor{blue}{49} \\ \textcolor{red}{0}} & 4, 3\tiny \makecell{\textcolor{blue}{10} \\ \textcolor{red}{0}}\\
\midrule
Java &  4 &  18 &  19 &  0 & 123, 108\tiny \makecell{\textcolor{blue}{191} \\ \textcolor{red}{87}} & 2, 2\tiny \makecell{\textcolor{blue}{3} \\ \textcolor{red}{2}} & 0, 0\tiny \makecell{\textcolor{blue}{0} \\ \textcolor{red}{0}} & 31, 22\tiny \makecell{\textcolor{blue}{73} \\ \textcolor{red}{12}} & 1, 0\tiny \makecell{\textcolor{blue}{7} \\ \textcolor{red}{0}} & 0, 0\tiny \makecell{\textcolor{blue}{4} \\ \textcolor{red}{0}} & 32,  13\tiny \makecell{\textcolor{blue}{246} \\ \textcolor{red}{4}} & 2, 1\tiny \makecell{\textcolor{blue}{7} \\ \textcolor{red}{0}} & 1, 0\tiny \makecell{\textcolor{blue}{4} \\ \textcolor{red}{0}} & 0, 0\tiny \makecell{\textcolor{blue}{0} \\ \textcolor{red}{0}}\\
\midrule
C/C++ &  35 &  2 &  28 &  10 & 133, 98\tiny \makecell{\textcolor{blue}{503} \\ \textcolor{red}{17}} & 3, 3\tiny \makecell{\textcolor{blue}{5} \\ \textcolor{red}{1}} & 3, 4\tiny \makecell{\textcolor{blue}{6} \\ \textcolor{red}{0}} & 102, 102\tiny \makecell{\textcolor{blue}{102} \\ \textcolor{red}{102}} & 5, 5\tiny \makecell{\textcolor{blue}{5} \\ \textcolor{red}{5}} & 9, 9\tiny \makecell{\textcolor{blue}{9} \\ \textcolor{red}{9}} & 44, 43\tiny \makecell{\textcolor{blue}{109} \\ \textcolor{red}{7}} & 4, 3\tiny \makecell{\textcolor{blue}{10} \\ \textcolor{red}{0}} & 4, 4\tiny \makecell{\textcolor{blue}{13} \\ \textcolor{red}{0}} & 41, 10\tiny \makecell{\textcolor{blue}{170} \\ \textcolor{red}{5}} \\
\midrule
JavaScript &  134  & 0  & 18  & 0  & 24, 10\tiny \makecell{\textcolor{blue}{420} \\ \textcolor{red}{5}} & 1, 1\tiny \makecell{\textcolor{blue}{6} \\ \textcolor{red}{0}} & 1, 0\tiny \makecell{\textcolor{blue}{25} \\ \textcolor{red}{0}} & 0, 0\tiny \makecell{\textcolor{blue}{0} \\ \textcolor{red}{0}} & 0, 0\tiny \makecell{\textcolor{blue}{0} \\ \textcolor{red}{0}} & 0,  0\tiny \makecell{\textcolor{blue}{0} \\ \textcolor{red}{0}} & 24, 23\tiny \makecell{\textcolor{blue}{45} \\ \textcolor{red}{4}} & 2,  2\tiny \makecell{\textcolor{blue}{4} \\ \textcolor{red}{2}} & 2, 2\tiny \makecell{\textcolor{blue}{3} \\ \textcolor{red}{1}} & 0, 0\tiny \makecell{\textcolor{blue}{0} \\ \textcolor{red}{0}} \\
\midrule
TypeScript &  230  & 0 &  9 &  9 & 32, 10\tiny \makecell{\textcolor{blue}{1841} \\ \textcolor{red}{5}} & 2, 2\tiny \makecell{\textcolor{blue}{4} \\ \textcolor{red}{1}} & 2, 2\tiny \makecell{\textcolor{blue}{5} \\ \textcolor{red}{0}} & 0, 0\tiny \makecell{\textcolor{blue}{0} \\ \textcolor{red}{0}} & 0, 0\tiny \makecell{\textcolor{blue}{0} \\ \textcolor{red}{0}} & 0, 0\tiny \makecell{\textcolor{blue}{0} \\ \textcolor{red}{0}} & 24, 21\tiny \makecell{\textcolor{blue}{34} \\ \textcolor{red}{12}} & 2, 2\tiny \makecell{\textcolor{blue}{3} \\ \textcolor{red}{1}} & 1, 1\tiny \makecell{\textcolor{blue}{1} \\ \textcolor{red}{0}} & 39, 33\tiny \makecell{\textcolor{blue}{144} \\ \textcolor{red}{1}}   \\

\bottomrule
\end{tabular}
\vspace{-0.3cm}
\end{table*}

\begin{figure*}[!t]
    \centering
    \includegraphics[width=1\textwidth]{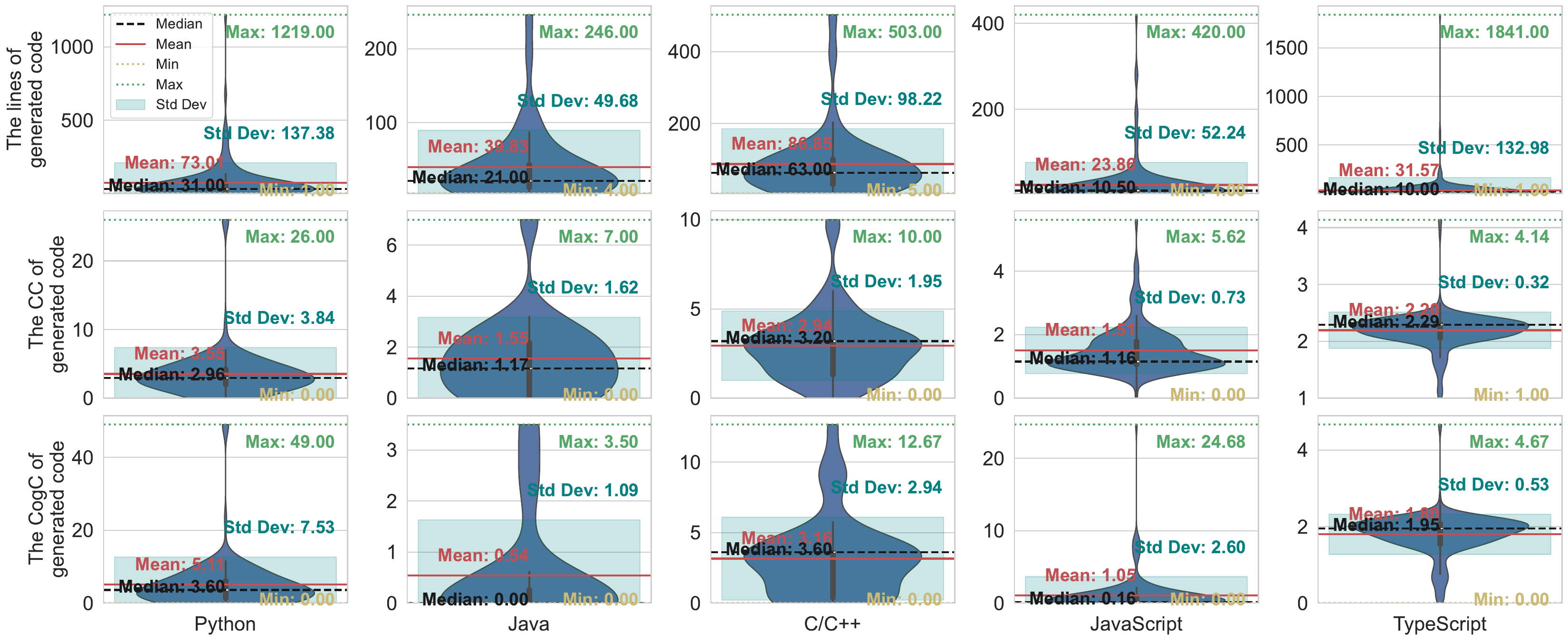}
    \vspace{-0.7cm} 
\caption{The violin plot distribution of the LOC, cyclomatic complexity, and cognitive complexity of the GPT-generated code.}
\label{fig:code_metrics}
\vspace{-0.5cm} 
\end{figure*}

Table \ref{tab:codetype} outlines the types of the self-admitted GPT-generated code across five programming languages. For all languages except Python, developers predominantly use ChatGPT and Copilot to create code scripts for algorithms and data structures, specifically utilizing common data structures for processing non-text data, along with fundamental algorithms. For example, the self-admitted GPT-generated C++ code \cite{PixelPilot_rk} implements algorithms to calculate and manage the average, rate, minimum, and maximum values within a specific time window. In the case of Python, text processing is also a dominant implementation type, which can be attributed to Python’s widespread use in data processing scenarios. For instance, the self-admitted GPT-generated Python code \cite{ecdo} reads an input file, removes newline characters, and writes the processed content to an output file.


Table \ref{tab:code} displays the granularity of the self-admitted GPT-generated code, including the number of files, classes, methods, and statements, along with the average, median, maximum, and minimum values of LOC, cyclomatic complexity (CC), and cognitive complexity (CogC) for these elements. Due to the majority of generated statements having a complexity of zero, the complexities of these statements are not listed. Figure \ref{fig:code_metrics} shows the distribution of LOC, cyclomatic complexity, and cognitive complexity for all self-admitted GPT-generated code. The median LOC for all code ranges from 10 to 63 across the five languages, with the majority of code snippets being relatively short and a smaller portion exhibiting moderate length. This suggests that the self-admitted GPT-generated code segments are typically concise, potentially reflecting their use for simpler tasks or as supplementary components within larger projects. For the longest code snippets in these five languages, Python is an exception where developers use ChatGPT to create a file that implements a module of mathematical computation tool functions ~\cite{math_tools}; 
for the other languages, the longest snippets ~\cite{TMoreTest.js, OpenApiV30.ts, DefaultIconPack.java, DovesLapTimer.cpp} typically result from developers using ChatGPT to mass-produce data or code with a fixed format.
Regarding complexity, the GPT-generated code exhibits low complexity levels, with median cyclomatic complexity values from 1 to 3 and cognitive complexity values between 0 and 4 across the five programming languages. This indicates that while developers use ChatGPT/Copilot to generate code of a certain scale, the output is generally simpler and less sophisticated, suitable for routine or boilerplate code and straightforward applications, but possibly inadequate for complex programming challenges.

Notably, C/C++ files and classes generated by ChatGPT/Copilot have higher LOC and complexity metrics than other languages. This reflects the developers' use of ChatGPT/Copilot to implement algorithms and data structures in C/C++, which typically demands more detailed and complex coding patterns.
In addition, TypeScript code statements generated by ChatGPT/Copilot tend to have higher LOC. This is mainly because TypeScript, with its enhanced type definitions, allows developers to use ChatGPT/Copilot to generate large batches of formatted data to perform exhaustive styling tests. This is often used to verify the display effects or correctness in frontend projects, such as with the ``textifai'' project
~\cite{textifai-color.ts}
. In this project, developers used ChatGPT to generate 84 colors for the interface to customize the styling.

\begin{center}
\vspace{-15pt}
    \resizebox{\linewidth}{!}{
\begin{tabular}{l!{\vrule width 1pt}p{0.95\columnwidth}}
    \makecell{{\LARGE \faLightbulbO}}  &\textbf{Finding 2.} 
     Developers primarily use ChatGPT and Copilot to generate code for algorithms and data structures across all five programming languages. In the case of Python, the self-admitted GPT-generated code for text processing is also common. 
     Developers commonly leverage ChatGPT/Copilot to generate code snippets that are either short or moderately sized, yet rarely overly complex.
\end{tabular}}
\vspace{-10pt}
\end{center} 

\begin{table*} [!ht]
\centering
 \caption{The code change information of the files containing the self-admitted GPT-generated code. \vspace{-0.2cm}
     }
    \label{tab:code-chang-info}
\fontsize{7.2pt}{8pt}\selectfont
\begin{tabular}{@{}l | c | c | l | l | l | c| l | l | l | c | c |  l@{}}
\toprule
 ~ & \#Files$_G$ & \#Files$_C$ & Stars$_C$ & Files$_C$ & Commits$_C$ & \#Change & Change & LOC$_{add}$ & LOC$_{delete}$ & \#Files$_B$ & \#Change$_B$ & Change$_B$  \\
\midrule
Python &  136  & 37 
& 176, 15\tiny \makecell{\textcolor{blue}{2.0K} \\ \textcolor{red}{5}}
& 1, 1\tiny \makecell{\textcolor{blue}{4} \\ \textcolor{red}{1}}
& 511, 137\tiny \makecell{\textcolor{blue}{2.9K} \\ \textcolor{red}{2}}
& 56  & 2, 1\tiny \makecell{\textcolor{blue}{5} \\ \textcolor{red}{1}}  & 19, 0\tiny \makecell{\textcolor{blue}{257} \\ \textcolor{red}{0}} & 9, 0\tiny \makecell{\textcolor{blue}{129} \\ \textcolor{red}{0}} & 8 & 11 & 1, 1\tiny \makecell{\textcolor{blue}{2} \\ \textcolor{red}{1}}  \\
\midrule

Java & 41  &  6 
& 43, 47\tiny \makecell{\textcolor{blue}{60} \\ \textcolor{red}{20}}
& 59, 1\tiny \makecell{\textcolor{blue}{348} \\ \textcolor{red}{1}}
& 206, 250\tiny \makecell{\textcolor{blue}{332} \\ \textcolor{red}{15}}
& 13  & 2, 2\tiny \makecell{\textcolor{blue}{4} \\ \textcolor{red}{1}}   & 5, 0\tiny \makecell{\textcolor{blue}{18} \\ \textcolor{red}{0}} &0, 0\tiny \makecell{\textcolor{blue}{0} \\ \textcolor{red}{0}} & 2 & 2 & 1, 1\tiny \makecell{\textcolor{blue}{1} \\ \textcolor{red}{1}}  \\
\midrule

C/C++ &  75 & 29 
& 58, 8\tiny \makecell{\textcolor{blue}{728} \\ \textcolor{red}{5}}
& 434, 1\tiny \makecell{\textcolor{blue}{3.4K} \\ \textcolor{red}{1}}
& 677, 67\tiny \makecell{\textcolor{blue}{2.8K} \\ \textcolor{red}{10}}
& 249 & 9, 2\tiny \makecell{\textcolor{blue}{38} \\ \textcolor{red}{1}}    & 24, 4\tiny \makecell{\textcolor{blue}{143} \\ \textcolor{red}{0}} & 3, 0\tiny \makecell{\textcolor{blue}{21} \\ \textcolor{red}{0}} & 9 & 19 & 2, 1\tiny \makecell{\textcolor{blue}{4} \\ \textcolor{red}{1}}\\
\midrule

JavaScript & 152  & 20 
& 574, 29\tiny \makecell{\textcolor{blue}{10.4K} \\ \textcolor{red}{5}}
& 48, 1\tiny \makecell{\textcolor{blue}{941} \\ \textcolor{red}{1}}
& 217, 92\tiny \makecell{\textcolor{blue}{812} \\ \textcolor{red}{7}}
& 51 & 3, 1\tiny \makecell{\textcolor{blue}{23} \\ \textcolor{red}{1}}  & 4, 0\tiny \makecell{\textcolor{blue}{20} \\ \textcolor{red}{0}} &10, 0\tiny \makecell{\textcolor{blue}{129} \\ \textcolor{red}{0}} & 6 & 13 & 2, 1\tiny \makecell{\textcolor{blue}{7} \\ \textcolor{red}{1}}  \\
\midrule

TypeScript & 248 &  19 
& 265, 187\tiny \makecell{\textcolor{blue}{2.0K} \\ \textcolor{red}{5}}
& 5, 1\tiny \makecell{\textcolor{blue}{79} \\ \textcolor{red}{1}}
& 683, 203\tiny \makecell{\textcolor{blue}{7.3K} \\ \textcolor{red}{48}}
& 31 & 2, 1\tiny \makecell{\textcolor{blue}{39} \\ \textcolor{red}{1}}  & 36, 1\tiny \makecell{\textcolor{blue}{1.0K} \\ \textcolor{red}{0}} &0, 0\tiny \makecell{\textcolor{blue}{0} \\ \textcolor{red}{0}} & 9 & 11 & 1, 1\tiny \makecell{\textcolor{blue}{2} \\ \textcolor{red}{1}}\\

\bottomrule
\end{tabular}
\vspace{-0.5cm}
\end{table*}

\subsection{RQ3: The characteristics of the code changes made to the self-admitted GPT-generated code}
\begin{table} [!t]
\centering
 \caption{The types of bugs leading to bug-fix code changes (The most frequent three types are highlighted in bold, and the percentages representing different types of bugs leading to bug-fix code changes may not total 100\% because the code could contain multiple types of bugs simultaneously.)  \vspace{-0.2cm}}

    \label{tab:bugfixtype}
\resizebox{\linewidth}{!}{
\begin{tabular}{lllllll}
\toprule
 ~ &  Type1  & Type2 & Type3 & Type4 & Type5 & Type6\\
\midrule
Python& \textbf{46\%}&  \textbf{30\%} & \textbf{35\%} & 4\% & 0 & 2\% \\
Java &  \textbf{50\%}& 0 &  \textbf{50\% }& 0 & 0 & 0 \\

C/C++ &  5\% & \textbf{35\%} & \textbf{55\%}& 0 & \textbf{10\%} & 5\%  \\

JS & 16\% & \textbf{20\%} & \textbf{48\%}& \textbf{20\%} & 0 & \textbf{24\%} \\

TS & 21\% & \textbf{43\%} & \textbf{46\%}& 18\% & 0 & 4\% \\
\midrule

Average & \textbf{28\%}& \textbf{26\%}& \textbf{47\%}& 8\%& 2\%& 10\%\\
\bottomrule
\end{tabular}}

\begin{tablenotes}[flushleft]
\item \textbf{Type Explanation:} Type1 (Interface and dependency management issues: problems related to incorrect API calls or incorrect recommended libraries), Type2 (Grammar and semantic errors: errors in the syntax or meaning of the code), Type3 (Algorithm logic errors: mistakes in the logic or flow of the implemented functions), Type4 (Performance issues: deficiencies affecting the speed or efficiency of the code), Type5 (Memory management issues: problems with the allocation, use, or release of memory), and Type6 (Security vulnerabilities: weaknesses that could be exploited to compromise the system). 
\end{tablenotes}
\vspace{-0.3cm}
\end{table}

\begin{table} [!t]
\centering
 \caption{The types of the non-bug-fix code change (The most frequent three types are highlighted in bold).\vspace{-0.2cm}
     }
    \label{tab:nonbugfixtype}
\resizebox{\linewidth}{!}{
\begin{tabular}{lllllll}
\toprule
 ~ &  Type1 &Type2 & Type3 & Type4 & Type5 & Others \\
\midrule
Python & \textbf{30\%}&   \textbf{24\%}&   19\%&  \textbf{20\%}& 2\% & 4\%\\
Java &   \textbf{50\%}&   \textbf{13\%}& \textbf{33\%} & 11\% &  0& 0 \\
C/C++ &   \textbf{23\%}&   \textbf{19\%}& \textbf{35\%} & 18\% &  3\% & 2\% \\
JavaScript&  \textbf{26\%} &   \textbf{22\%}& 17\%&  \textbf{20\%}&    4\%& 10\%\\
TypeScript&  \textbf{55\%}&   6\%&  \textbf{14\%}& \textbf{14\%}&  0& 10\%\\
\midrule
Average & \textbf{37\%}&  \textbf{17\%}& \textbf{24\%}& 17\%& 2\%& 5\%\\
\bottomrule

\end{tabular}}

\begin{tablenotes}[flushleft]
\item \textbf{Type Explanation:} Type 1 (Feature addition), Type 2 (Optimization: optimizing existing code to enhance performance and reduce resource consumption), Type 3 (Refactoring), Type 4 (Style adjustment: adjusting the code's formatting, indentation, and naming conventions.), Type 5 (Removal of unused code), and Others (e.g., documentation comments update, etc.).
\end{tablenotes}
\vspace{-0.3cm}
\end{table}

Table \ref{tab:code-chang-info} displays the code change information for the self-admitted GPT-generated code. 
``\#Files$_G$'' is the number of files containing the self-admitted GPT-generated code. ``\#Files$_C$'' is the number of files where the self-admitted GPT-generated code is changed. The ``Star$_{C}$'', ``File$_{C}$'', and ``Commits$_{C}$'' column entries represent the average, median, maximum, and minimum values of the stars, files, and commits of the projects where the self-admitted GPT-generated code is modified. 
``\#Change'' indicates the number of times that the self-admitted GPT-generated code is modified in these files. 
The ``LOC$_{add}$'' column entries represent the average, median, maximum, and minimum values of LOC added across all code changes, while the ``LOC$_{delete}$'' column entries show these values for  LOC deleted. 
``\#Files$_B$'' shows the number of files where the self-admitted GPT-generated code is modified due to bug fixes, and ``\#Change$_B$'' represents the number of times that the self-admitted GPT-generated code is modified due to bug fixes in these files. 
The ``Change'' column entries represent the average, median, maximum, and minimum values for the number of times the self-admitted GPT-generated code is modified, while the ``Change$_B$'' column entries show these values for modifications due to bug fixes.

Among the self-admitted GPT-generated code, 39\% C/C++ code has been modified, followed by 27\% for Python, 15\% for Java, 13\% for JavaScript, and 8\% for TypeScript. The median number of modifications to the self-admitted GPT-generated code ranges between 1 and 2 across these five languages. For all code changes, the median number of LOC added is between 0 and 4, while the median number of LOC deleted is 0. This suggests that the self-admitted GPT-generated code typically undergoes few modifications once created, and those that do occur are generally minor, reflecting targeted adjustments rather than extensive rewrites or overhauls. However, there are rare instances involving substantial modifications to the self-admitted GPT-generated code. For example, the GPT-generated code in the ``vaporview'' project \cite{vaporview} has undergone 39 changes from the initial commit \cite{vaporview-commit-f35d} to the latest modification \cite{vaporview-commit-a50c}, involving various types of modifications such as feature additions, bug fixes, optimizations, and refactoring. These changes ultimately increase the total LOC by 1,031.

12\% of the self-admitted GPT-generated C/C++ code is modified due to bugs, with Python at 6\%, Java at 5\%, JavaScript at 4\%, and TypeScript at 4\%. 
The median number of bug-fix modifications to GPT-generated code is 1 across the five programming languages, indicating that bug-related modifications in GPT-generated code are relatively infrequent. 




\begin{center}
\vspace{-15pt}
    \resizebox{\linewidth}{!}{
\begin{tabular}{l!{\vrule width 1pt}p{0.95\columnwidth}}
    \makecell{{\LARGE \faLightbulbO}}  &\textbf{Finding 3.} 
   GPT-generated code typically undergoes few modifications after its creation, often necessitating only incremental adjustments rather than extensive overhauls. Additionally, the infrequency of bug-related changes suggests that ChatGPT/Copilot tends to produce code of relatively high initial quality.
\end{tabular}}
\vspace{-10pt}
\end{center}

\begin{figure}[!t]
            \includegraphics[width=1\linewidth]{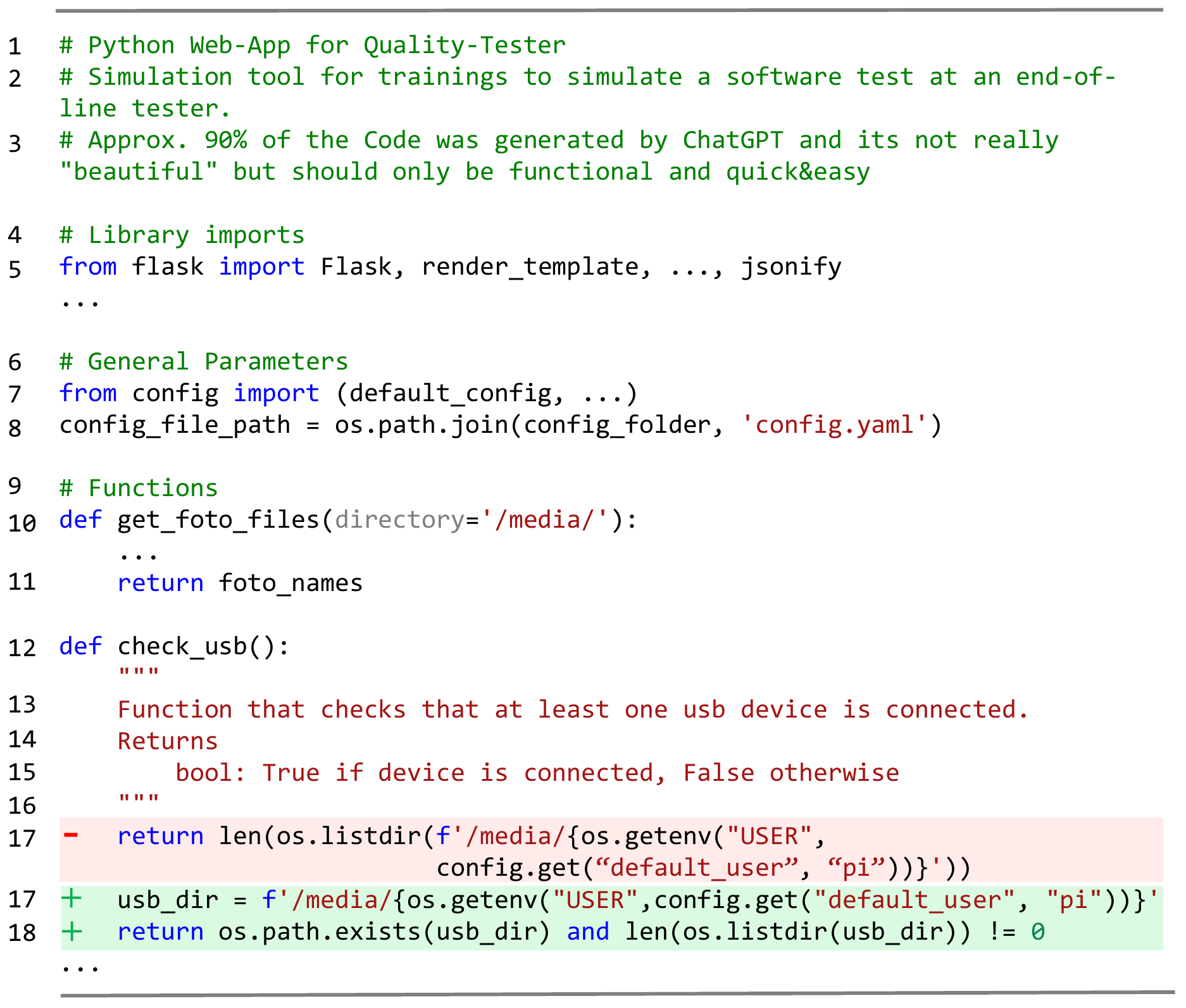}             
            \vspace{-0.6cm} 
        \caption{ \vspace{-0.1cm}A bug-fix modification made to the GPT-generated code, which contains an algorithmic logic error. \vspace{-0.2cm} }
        \label{fig:bug-fix-change}
        \vspace{-0.5cm}
\end{figure}

Table \ref{tab:bugfixtype} presents the types of bugs leading to bug-fix code changes.  The most common types of bugs are algorithmic and logic errors, interface and dependency management issues, and grammar and semantic errors. Additionally, we discover that among 56 code snippets with bug-related modifications, 6 have comments indicating that most of the code is generated by ChatGPT, and 3 are flagged as potentially buggy or requiring testing. This suggests these snippets may involve complex tasks that ChatGPT struggles to handle, requiring developers to modify or add human-written code. Additionally, when uploaded to GitHub, this code might not undergo thorough testing, making it more susceptible to bugs. For example, in the ``quality-tester'' project, the code comment for the GPT-generated file ``main.py'' ~\cite{quality-tester-commit-3c14}
shown in Figure \ref{fig:bug-fix-change} is: ``\textit{\# Approx. 90\% of the code was generated by ChatGPT, and it is not really ‘beautiful' but should only be functional and quick \& easy.}''
The GPT-generated code shown in Figure \ref{fig:bug-fix-change} contains an algorithmic logic error, where the ``\texttt{check\_usb}'' function in lines 12-17 directly uses ``\texttt{os.listdir}'' to get a directory list without verifying the directory's existence. 
In a subsequent code commit ~\cite{quality-tester-commit-58bd}, a bug fix is applied to this code to correct the algorithmic logic error (commit message: ``\textit{Fixed bug that tester did not execute when no USB device was ever connected.}''). The fix, shown in lines 17-18 of Figure \ref{fig:bug-fix-change}, involves adding an ``\texttt{os.path.exists(usb\_dir)}'' check in the original ``\texttt{check\_usb}'' function, ensuring that the directory exists before attempting to list its contents.

\begin{figure}[!t]
        \includegraphics[width=\linewidth]{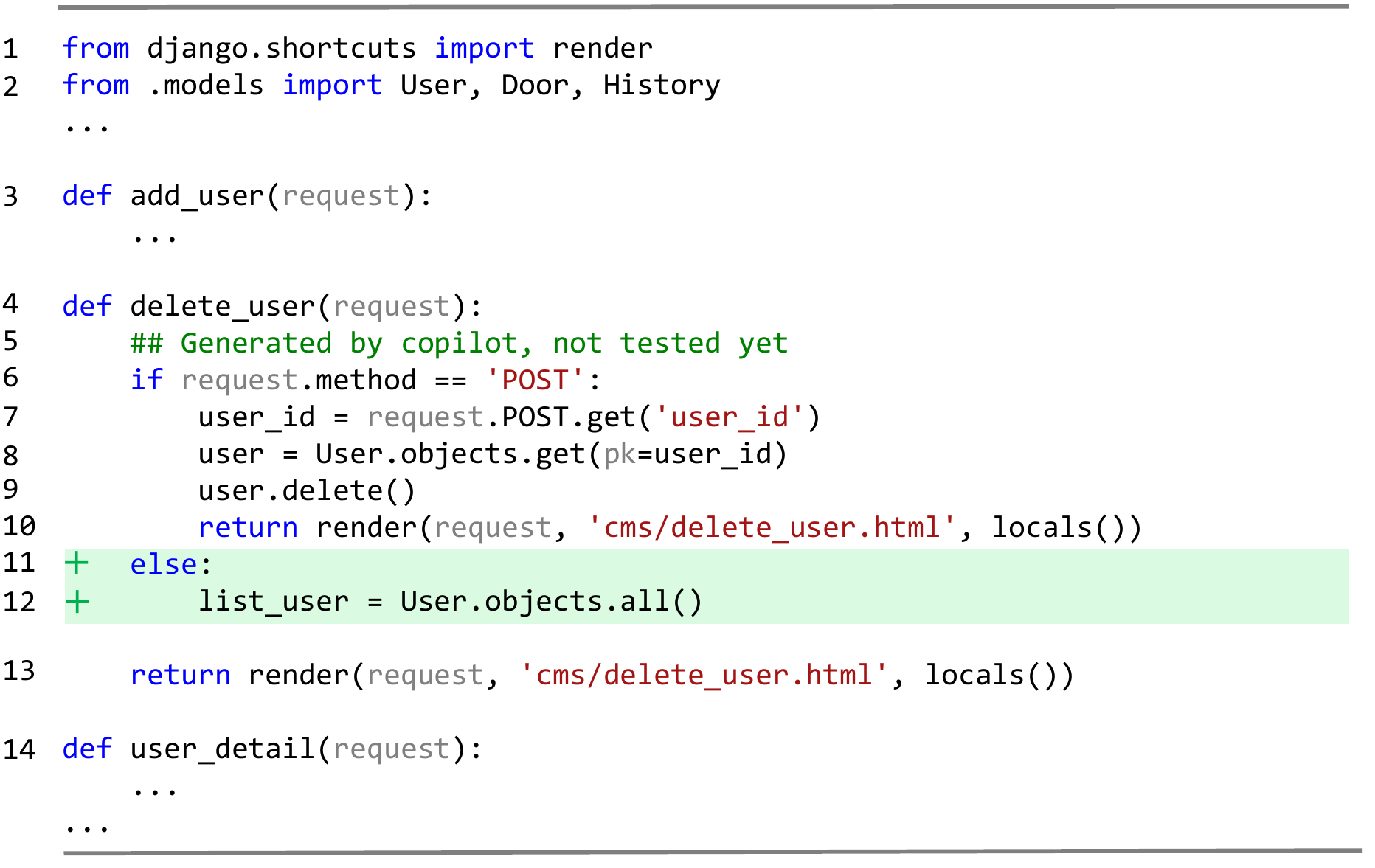}
        \vspace{-0.6cm} 
        \caption{A feature addition made to the GPT-generated code.\vspace{-0.2cm}}
        \label{fig:feature-addition}
       \vspace{-0.5cm} 
\end{figure}

Table \ref{tab:nonbugfixtype} displays the various types of non-bug-fix code modifications for the self-admitted GPT-generated code. The most common modification, feature addition, entails introducing new features or functionalities to address emerging project needs or to enhance current features. Refactoring, which is the practice of reorganizing existing code to enhance its maintainability and scalability without changing its functionality, ranks as the second most frequent alteration.  
For example, Figure \ref{fig:feature-addition} shows that the ``\texttt{delete\_user}''
~\cite{access_control_web-commit-79b8}
function is generated by Copilot. The comment for the generated code is ``\textit{\#\# Generated by copilot, not tested yet}''. 
The generated code in Figure \ref{fig:feature-addition} allows the deletion of a user when a POST request is made with the user's ID.
In a subsequent code commit 
~\cite{access_control_web-commit-ddef}
, a new feature is added to handle non-POST requests and display a list of all users. The feature addition, shown in lines 11-12 of Figure \ref{fig:feature-addition}, introduces new logic for non-POST requests by adding the functionality to retrieve a list of all users.

\begin{center}
\vspace{-15pt}
\resizebox{\linewidth}{!}{
\begin{tabular}{l!{\vrule width 1pt}p{0.95\columnwidth}}
\makecell{{\LARGE \faLightbulbO}} &\textbf{Finding 4.}
Algorithmic and logic errors, grammar and semantic errors, and interface and dependency management challenges are the primary types of bugs in GPT-generated code. The most common reasons for non-bug-fix code changes include feature addition and refactoring.  
\end{tabular}}
\vspace{-10pt}
\end{center}

\subsection{RQ4: The difference between the self-admitted GPT-generated code and code without explicit GPT-generation annotation}
\begin{table}[!t]
\centering
\caption{The ratio of LOC, average method cyclomatic complexity (CC$m$), average method cognitive complexity (CogC$m$), average modification (Change), and average bug-fix modification (Change$B$) for the self-admitted GPT-generated code in the projects.\vspace{-0.2cm}}
\label{tab:ratio}
\fontsize{6.5pt}{8pt}\selectfont
\begin{tabular}{@{}p{1.0cm} | c | c | c | c | c l@{}}
\toprule

  ~& LOC & CC$_m$ & CogC$_m$ &  Change  & Change$_B$ \\ 

\midrule
Python & 0.16, 0.03\tiny \makecell{\textcolor{blue}{1.0} \\ \textcolor{red}{0}} & 1.01, 0.79\tiny \makecell{\textcolor{blue}{10.5} \\ \textcolor{red}{0.4}} & 1.67, 0.53\tiny \makecell{\textcolor{blue}{27.5} \\ \textcolor{red}{0.1}} & 0.16, 0\tiny \makecell{\textcolor{blue}{4057} \\ \textcolor{red}{0}} & 0.27, 0\tiny \makecell{\textcolor{blue}{56.3} \\ \textcolor{red}{0}} \\ \midrule
Java & 0.07, 0.02\tiny \makecell{\textcolor{blue}{0.9} \\ \textcolor{red}{0}} & 1.52, 1.41\tiny \makecell{\textcolor{blue}{5.9} \\ \textcolor{red}{0.4}} &2.40, 1.50\tiny \makecell{\textcolor{blue}{18.5} \\ \textcolor{red}{0.2}} & 0.16, 0\tiny \makecell{\textcolor{blue}{6.0} \\ \textcolor{red}{0}} & 0.40, 0\tiny \makecell{\textcolor{blue}{21.0} \\ \textcolor{red}{0}} \\ \midrule
C/C++ & 0.08, 0.03\tiny \makecell{\textcolor{blue}{0.5} \\ \textcolor{red}{0}} & 1.34, 1.13\tiny \makecell{\textcolor{blue}{7.4} \\ \textcolor{red}{0.3}} &1.80, 1.08\tiny \makecell{\textcolor{blue}{13.1} \\ \textcolor{red}{0.2}} & 2.70, 0\tiny \makecell{\textcolor{blue}{76.6} \\ \textcolor{red}{0}} &0.60, 0\tiny \makecell{\textcolor{blue}{19.3} \\ \textcolor{red}{0}}  \\ \midrule
JavaScript & 0.11, 0.02\tiny \makecell{\textcolor{blue}{0.8} \\ \textcolor{red}{0}} & 1.29, 0.93\tiny \makecell{\textcolor{blue}{5.1} \\ \textcolor{red}{0.2}} &2.10, 0.78\tiny \makecell{\textcolor{blue}{73.9} \\ \textcolor{red}{0.3}} & 1.70, 0\tiny \makecell{\textcolor{blue}{25.4} \\ \textcolor{red}{0}} &0.11, 0\tiny \makecell{\textcolor{blue}{10.3} \\ \textcolor{red}{0}}  \\ \midrule
TypeScript & 0.09, 0.03\tiny \makecell{\textcolor{blue}{0.7} \\ \textcolor{red}{0.1}} & 1.42, 1.22\tiny \makecell{\textcolor{blue}{5.5} \\ \textcolor{red}{0.2}} &2.09, 1.16\tiny \makecell{\textcolor{blue}{22.4} \\ \textcolor{red}{0.2}} & 0.21, 0\tiny \makecell{\textcolor{blue}{7.5} \\ \textcolor{red}{0}} &0.09, 0\tiny \makecell{\textcolor{blue}{3.3} \\ \textcolor{red}{0}}  \\
\bottomrule
\end{tabular}
\vspace{-0.5cm} 
\end{table}

The first column of Table \ref{tab:ratio} shows the statistic of the proportion of the self-admitted GPT-generated LOC to the total LOC in projects. The median proportion is notably low, ranging from 0.02 to 0.03, signifying that the self-admitted GPT-generated code typically forms a minor part of the projects. However, the maximum proportion reaches up to 1.0 in certain cases. This is observed when the initial commit introducing the self-admitted GPT-generated code consists exclusively of such code, without any human-written content, such as the ``deovr-remote-multihead'' project \cite{deovr}.

The second and third columns of Table \ref{tab:ratio} illustrate the ratios of average method Cyclomatic Complexity and Cognitive Complexity between the GPT-generated and code without explicit GPT-generation
annotation. The calculation method is shown in equation (\ref{eq:ratio1}) and (\ref{eq:ratio2}), respectively. 

\vspace{-0.1cm}
\begin{equation}
\frac{CC_\text{GPT$_M$} / \#M_\text{GPT}}{ \left(CC_\text{Total} - CC_\text{GPT$_M$}\right) /\left( \#M_\text{Total} - \#M_\text{GPT}\right)},
\label{eq:ratio1}
\end{equation}
\vspace{-0.3cm}
\begin{equation}
\frac{CogC_\text{GPT$_M$} / \#M_\text{GPT}}{ \left(CogC_\text{Total} - CogC_\text{GPT$_M$}\right) /\left( \#M_\text{Total} - \#M_\text{GPT}\right)},
\label{eq:ratio2}
\end{equation}
\vspace{-0.4cm}

\noindent where $CC_\text{GPT$_M$}$ represents the total CC of GPT-generated methods, $\#M_\text{GPT}$ denotes the number of GPT-generated methods, $CC_\text{Total}$ is the total CC across all project methods, and $\#M_\text{Total}$ is the total number of project methods. Similarly, $CogC_\text{GPT$_M$}$ stands for the CogC of the self-admitted GPT-generated methods, while $CogC_\text{Total}$ is the total cognitive complexity across all project methods. For Python and JavaScript, the median ratios are below 1, indicating that the self-admitted GPT-generated code has lower CC and CogC than code without explicit GPT-generation
annotation in most projects. Conversely, for C/C++, Java, and TypeScript, the median values exceed 1, suggesting higher complexities in the self-admitted GPT-generated code compared to code without explicit GPT-generation
annotation in these languages. This inconsistency in complexity ratios across different languages might be attributed to the diverse nature of code without explicit GPT-generation
annotation in various projects. Due to the labor-intensive of analyzing the code types, this aspect is not explored in our current research but is identified as an area for future studies. 

The fourth and fifth columns of Table \ref{tab:ratio} display the ratios of average modifications and bug-fix modifications for the self-admitted GPT-generated code compared to all code. The calculation method is shown in equations (\ref{eq:ratio3}) and (\ref{eq:ratio4}).

\vspace{-0.1cm}
\begin{equation}
\frac{\#Commits_\text{GPT} / \#Files_\text{GPT}}{\#Commits_\text{Total} / \#Files_\text{Total}},
\label{eq:ratio3}
\end{equation}
\vspace{-0.3cm}
\begin{equation}
\frac{\#BugCommits_\text{GPT} / \#Files_\text{GPT}}{\#BugCommits_\text{Total} / \#Files_\text{Total}}, 
\label{eq:ratio4}
\end{equation}
\vspace{-0.4cm}

\noindent where $\#Commits_\text{GPT}$ represents the number of commits to GPT-generated code, $\#Files_\text{GPT}$ is the number of the GPT-generated code, $\#Commits_\text{Total}$ is the total number of project commits, and $\#Files_\text{Total}$ is the total number of files in the project. Similarly, $\#BugCommits_\text{GPT}$ represents the number of bug-fix commits to GPT-generated code, and $\#BugCommits_\text{Total}$ is the total number of bug-fix commits in the project. The median ratio is close to 0, indicating that GPT-generated code undergoes fewer modifications and bug fixes relative to code without explicit GPT-generation
annotation. This phenomenon may primarily stem from two factors: first, GPT-generated code comprises only a small portion of the overall project, so modifications or new features typically focus on code without explicit GPT-generation
annotation; second, the code generated by ChatGPT/Copilot usually involves non-core business logic, such as data processing and transformation scripts, which generally meet initial requirements and seldom require further modifications. While there are outliers with exceptionally high ratios, the overall trend is low modification rates.  For instance, in the ``AIGenerated'' project \cite{aigenerated}
, a single GPT-generated file is modified once out of two total commits in a project with 8114 files, resulting in a ratio of 4057 (=(1/1)/(2/8114)).

\begin{center}
\vspace{-15pt}
\resizebox{\linewidth}{!}{
\begin{tabular}{l!{\vrule width 1pt}p{0.95\columnwidth}}
\makecell{{\LARGE \faLightbulbO}} &\textbf{Finding 5.}
ChatGPT-generated code constitutes only a small portion of all projects. Compared to human-written code, ChatGPT-generated code rarely undergoes changes or bug-fix modifications.
\end{tabular}}
\vspace{-10pt}
\end{center}

\subsection{RQ5: The annotation of the self-admitted GPT-generated code}

We investigate the information contained in the code comment that indicates the code snippet is generated by ChatGPT/Copilot. We discover six primary types of information, as depicted in Table \ref{tab:comment}: the version of LLMs employed (Ver), the timestamp of code generation (Time), the prompt provided to ChatGPT/Copilot for code generation (Prompt), the proportion of code generated by ChatGPT/Copilot in the snippet (Pro), whether the GPT-generated code undergoes human modifications (Modified), and whether the code necessitates further testing, review, or modifications, or has already been tested (Tested).  We select the following 12 code comments as examples to explain the six primary types of information. 

On average, 23\% of the code comments indicate the version of LLMs used (e.g., ChatGPT-3.5, ChatGPT-4, ChatGPT-4o, Copilot), as shown in comments 1, 3, 4, and 6. 9\% indicate the timestamp of code generation, exemplified by comment 2. 11\% show the prompt used, as seen in comments 4 and 5. 4\% indicate the proportion of GPT-generated code in the snippet, as shown by comments 1 and 10. 6\% indicate human modifications, demonstrated by comment 5. Finally, 3\% indicate whether the code requires further testing or has already been tested, as evidenced by comments 6-11. Additionally, many code comments simply indicate that the code is generated by ChatGPT/Copilot without providing any of the six types of information, such as comment 12.

\noindent\textbf{\textit{Comment 1:}} 
\textit{I wrote ``prompt'' comments and the whole thing was generated by Copilot.}

\noindent\textbf{\textit{Comment 2:}}  \textit{Generated by GPT4o (2024/05/27).}

\noindent\textbf{\textit{Comment 3:}}  \textit{This code was generated by ChatGPT 3.5. Prompt: write me a web scraper in Python that sends the text content of the webpage to the ChatGPT API, then outputs the email addresses on that webpage to a local text file.}

\noindent \textbf{\textit{Comment 4:}} \textit{ original test code generated by GPT-4: @see https://chat.openai.com\/share\/bcf2f05d-2c46-4920-8c8f-9b5d6a9536ff}

\noindent\textbf{\textit{Comment 5:}}\textit{Tests generated by ChatGPT, and modified by me}

\noindent\textbf{\textit{Comment 6:}}  \textit{below is generated by copilot, will probably rewrite these later on}

\noindent\textbf{\textit{Comment 7:}}  \textit{Created by ChatGPT TODO: Test the formula}

\noindent\textbf{\textit{Comment 8:}}  \textit{NOTE: This code was generated by ChatGPT, as I couldn't find other implementaions online, I take no credit for this code. }

\noindent\textbf{\textit{Comment 9:}}  \textit{! DANGER: this code is generated by chatGPT, and I have no idea what it did but it works.}

\noindent\textbf{\textit{Comment 10:}}  \textit{ Most of this code was generated by ChatGPT but was tested by co-authors to ensure its correctness }

\noindent\textbf{\textit{Comment 11:}}    \textit{This function was written by ChatGPT but have been tested and does the work.}

\noindent\textbf{\textit{Comment 12:}}    \textit{ This file was generated by ChatGPT}

\begin{table} [!t]
\centering
 \caption{The information contained in comments indicating that the code snippet is generated by ChatGPT/Copilot. 
     }
    \label{tab:comment}
\begin{tabular}{@{}l  l  c  c  l  c  c@{}}
\toprule
  & Ver & Time & Prompt & Pro & Modified  &Tested \\
\midrule
Python & 29\%&   3\%&    29\%&    9\%&   17\%&    5\%    \\
Java &  54\%&   15\%&    15\%&     5\%&   5\%&    5\%     \\
C/C++ &  19\%&   24\%&   12\%&    3\%&   3\%&    5\%     \\
JavaScript &    8\%&   0.7\%&    0.7\%&    0\%&    2\%&    0\%     \\
TypeScript &    4\%&   0.4\%&    0.4\%&    0.8\%&    2\%&    0.4\%    \\
\midrule
 Average &      23\%&   9\%&    11\%&    4\%&   6\%&    3\%\\
\bottomrule
\end{tabular}
\vspace{-0.5cm}
\end{table}

\begin{center}
\vspace{-10pt}
    \resizebox{\linewidth}{!}{
\begin{tabular}{l!{\vrule width 1pt}p{0.95\columnwidth}}
    \makecell{{\LARGE \faLightbulbO}}  &\textbf{Finding 6.}  A minority of code comments include details like the timestamp, the prompt provided, the proportion of GPT-generated code, human modifications, and further testing or review. However, most comments simply indicate that the code is generated by ChatGPT/Copilot.
\end{tabular}}
\vspace{-8pt}
\end{center}

\section{Discussion}
\label{sec: discussion}

\subsection{Implications}

\label{sec:implications}

Liang et al.~\cite{liang2024large} surveyed 410 developers and found that they may be discouraged from using LLMs to generate complex code due to the high likelihood of errors or the significant time required to debug or modify the generated code. \textbf{Our finding that the self-admitted GPT-generated code on GitHub tends to be relatively short and less complex validates Liang et al.'s conclusions.} 
Indeed, ChatGPT struggles with complex tasks, such as non-standalone function generation and class-level code generation, as evidenced by its performance in CoderEval~\cite{yu2024codereval} and ClassEval~\cite{du2024evaluating}, with a pass@1 rate of around 35\%. 
Despite this, GPT-generated code on GitHub shows a low bug incidence, likely because ChatGPT/Copilot excels at generating relatively short and straightforward code, such as data processing and transformation scripts. Thus, we \textbf{encourage developers to use ChatGPT/Copilot for generating code for less complex tasks}. 

The self-admitted GPT-generated code constitutes only a small portion of total projects and is not complex. Future studies could \textbf{investigate why most code in projects containing GPT-generated code is not created using LLMs and how to better utilize LLMs for complex programming tasks, e.g., class-level code generation}. Some LLMs, e.g., DeepSeek, show comparable performance to ChatGPT \cite{zhu2024deepseek}, yet very little self-admitted code generated by them has been found on GitHub. Subsequent research could \textbf{explore why LLMs like DeepSeek are underutilized and how to increase their adoption by developers}. 

Our collected GPT-generated code available in  ~\cite{SupplementalMaterials} offers valuable resources for subsequent research, particularly as \textbf{a testing dataset to develop and validate models for detecting LLM-generated code}. Researchers can integrate our curated collection of self-admitted LLM-generated code snippets with pre-LLM-era GitHub code (non-LLM-generated) into a unified benchmark dataset. This combined corpus enables rigorous evaluation of detection models ~\cite{pan2024assessing, shi2024between}, testing their ability to accurately distinguish LLM-generated code from human-written code. Additionally, our dataset supports constructing more realistic benchmarks for LLM code generation evaluation.  We acknowledge that programmers' daily development tasks certainly include handling complex programming tasks like those in ClassEval and CoderEval. However, developers currently do not tend to use LLMs to generate this complex code due to their low performance ~\cite{liang2024large}.  
Therefore, future researchers can \textbf{adapt the GPT-generated code to construct benchmarks to evaluate LLMs in scenarios where developers actually use LLMs for code generation}. 

In code review and maintenance\cite{abdelaziz2024integrating}—such as within GitHub’s CI/CD workflows—the key bug types identified in GPT-generated code (algorithmic and logic errors, grammar and semantic errors, and interface and dependency management challenges) and the common reasons for code changes (feature addition and refactoring) observed in our analysis make code reviews more targeted. \textbf{Reviewers can prioritize these high-risk areas when examining GPT-generated code, streamlining reviews by focusing on known vulnerabilities instead of broad, unfocused checks.}

Code comments indicating GPT-generated code encompass various details, such as  the prompt provided to ChatGPT/Copilot, the proportion of GPT-generated code, indications of human modifications, and whether the code requires further testing, review, or modifications, or has already been tested. However, there exists \textbf{a lack of standardized guidelines regarding when and how to apply code comments to GPT-generated code and what specific information should be included in these comments}. For instance, should the code generated by ChatGPT/Copilot still be labeled as such if it has already undergone testing? Furthermore, if time constraints prevent thorough testing of GPT-generated code, should additional information, such as the proportion of GPT-generated code and the provided prompt, be included in the comments to facilitate future testing efforts? Hence, future researchers may address these questions by \textbf{conducting surveys or interviews with developers and project managers to inform best practices for commenting on and managing the GPT-generated code in software development}.  

\subsection{Threats of Validity}
\label{sec: threats of validity}

\noindent \textbf{\textit{Search keywords}.} We construct GitHub search keywords like ``\textit{generated by ChatGPT}'' by analyzing common triplets containing the tokens such as ``ChatGPT'', ``Copilot'', ``GPT3'', ``GPT4'', ``GPT-3'', and ``GPT-4''. However, some comments may imply ChatGPT/Copilot usage without these keywords.
We also attempt searches with broader keywords like ``auto-generated'' and ``automated generated'' that do not specify the names of LLMs, but these searches do not yield significant results for code generated by LLMs. This indicates that most of code generation references are tied to specific model names. Additionally, we do not account for non-English comments, as English is an internationally recognized language.

\noindent \textbf{\textit{GPT-generated code}.} 
(a) Developers may use ChatGPT/Copilot to generate code and then make slight modifications to fit their projects ~\cite{grewal2024analyzing}. We do not differentiate between entirely generated and modified code. This does not significantly impact our analysis, as developers typically make minor modifications and often comment,  ``This file is mostly generated by ChatGPT.'' Additionally, some developers might not label the code as generated by LLMs in the comments. We have collected a total of 696 GPT-generated code snippets from projects with more than 5 stars, which can help clarify the characteristics of code generated by developers using LLMs in real-world software development scenarios to some extent. 
(b) We do not analyze the differences between code generated by ChatGPT and Copilot because both are based on GPT models. Copilot-generated code accounts for only about 10\% of all GPT-generated code.
Additionally, we observe that under comments like ``generated by Copilot'', the code often consists of entire methods, indicating that Copilot is frequently used for code generation rather than code completion.
(c) We are aware that some large closed-source companies, such as Microsoft (Copilot), Baidu (Comate), and Alibaba (TONGYI Lingma), integrate their LLMs into their IDEs and mainly use them for code completion rather than code generation in their business logic code. However, we cannot access the code from these closed-source companies, so our analysis is limited to the GPT-generated code available on GitHub. 

\section{Related Work}
\label{sec:related work}

\noindent \textbf{\textit{Quality analysis of code generated by LLMs.}} 
Some studies\cite{xia2024aicodereval,paul2024sceneval,zhuo2024bigcodebench,liu2024codeupdatearena,matton2024leakage,shao2024case2code, chen2021evaluating, austin2021program, yu2024codereval} are dedicated to proposing benchmarks for code generation with LLMs. For example, early research introduced benchmarks like HumanEval \cite{chen2021evaluating} and MBXP \cite{austin2021program}, which involve standalone function generation. 
Recently, Yu et al. \cite{yu2024codereval} proposed CoderEval for generating non-standalone functions that incorporated contextual dependencies and non-primitive types. 
Du et al. ~\cite{du2024evaluating} introduced ClassEval to tackle class-level code generation.  Some researchers ~\cite{liu2024empirical, jesse2023large, yeticstiren2023evaluating, yeo2024framework, song2023empirical, al2022readable, liu2023refining, siddiq2022empirical, liu2024no}
aim to explore more quality attributes of code generated by LLMs. For instance, Yeticstiren et al. ~\cite{yeticstiren2023evaluating} used SonarQube to evaluate the generated code of ChatGPT, Copilot, and CodeWhisperer in terms of code validity, code correctness, code security, code reliability, and code maintainability. In addition, some studies ~\cite{tambon2025bugs} have also explored the types of errors in code generated by LLMs. 
However, these existing studies primarily focus on assessing LLMs' code generation capabilities using controlled datasets \cite{chen2024survey}, which may not fully capture the characteristics of LLMs-generated code in actual software development scenarios \cite{wu2024versicode}. 


\noindent \textbf{\textit{Usage analysis of ChatGPT in code-related tasks.} } Recent studies have focused on the usage analysis of ChatGPT in code-related tasks.  Xiao et al. ~\cite{xiao2024devgpt} introduced the DevGPT dataset, which features conversations between developers and ChatGPT that are explicitly linked in GitHub projects. This dataset includes commits, pull requests, files, issues, and discussions that contain developer-ChatGPT interactions. Jin et al. \cite{jin2024can} utilized the DevGPT dataset to analyze how developers with ChatGPT for code generation and to assess the helpfulness of the code produced by ChatGPT in assisting developers. Chouchen et al. \cite{chouchen2024software} also leveraged the DevGPT dataset to compare pull requests that involve ChatGPT with those that do not. Das et al. \cite{das2024developers} analyzed 289 Developer-ChatGPT conversations to understand the usage of ChatGPT in GitHub issues. Watanabe et al. ~\cite{watanabe2024use} investigated 229 review comments to analyze how developers react to the information and suggestions provided by ChatGPT. Tufano et al. \cite{tufano2024unveiling} examined 165 commits, 159 pull requests, and 143 issues that included Developer-ChatGPT conversations, identifying 45 software engineering-related tasks that developers automate using ChatGPT. Grewal et al. \cite{grewal2024analyzing} analyzed the DevGPT dataset and found that a median of 54\% of lines of code from ChatGPT-generated snippets are embedded in GitHub projects. They noted that only a small number of code snippets changed in subsequent commits, with some being deleted, while the remaining snippets underwent modifications, often involving minor functionality changes, code reorganization, or name refinements. Our research shares similarities with Grewal et al.'s  \cite{grewal2024analyzing} work; however, we focus on directly analyzing GitHub projects where developers explicitly acknowledge the use of LLM-generated code. In our study, we select projects with star counts above 5 to avoid toy programs, ensuring that our analysis reflects real-world usage. We also examine the characteristics of projects containing ChatGPT-generated code, the characteristics of the generated code, and how developers annotate and label this code. Our findings provide a deeper understanding of the integration of LLM-generated code into software development workflows.

\section{Conclusion and Future Work}
\label{sec:conclusion}

We investigate the characteristics of the self-admitted GPT-generated code and associated projects on GitHub, revealing key insights. Most GitHub projects with GPT-generated code are not large but are dynamic and improving. 
Developers prefer using ChatGPT/Copilot for generating algorithms and data structures code for Python, Java, C/C++, JavaScript, and TypeScript. In Python, generating code for text processing is also common. The self-admitted GPT-generated code makes up a small portion of projects, undergoes fewer modifications, and has a lower incidence of bugs. Few code comments detail the prompts used, human modifications, and testing or review status. Based on these findings, we provide several implications. 
Future research will examine code smells and readability issues in GPT-generated code on GitHub, as our current focus has been on identifying bugs. 
To facilitate replication, the source code for scraping GPT-generated code, along with the self-admitted GPT-generated code and its characteristics, is available in our replication package ~\cite{SupplementalMaterials}.

\bibliographystyle{IEEEtran}
\bibliography{main.bib}

\end{document}